\documentclass[
reprint,
superscriptaddress,
amsmath,amssymb,
aps,
prl
]{revtex4-2}

\usepackage[english]{babel}
\usepackage{graphicx}
\usepackage[colorlinks=true, allcolors=blue]{hyperref}
\usepackage{booktabs}
\usepackage[normalem]{ulem}
\usepackage{mhchem}
\usepackage{pifont}
\usepackage{makecell}
\usepackage{gensymb}
\usepackage{braket}

\begin{document}
	
	\title{Medium-Throughput Evaluation of Quantum Geometry-Driven Topological Transports in Altermagnets}
	
	\author{Fu Li}
	\thanks{These authors contributed equally to this work.}
	\affiliation{Institute of Materials Science, Technical University of Darmstadt, Darmstadt 64287, Germany}
	
	\author{Bo Zhao}
	\thanks{These authors contributed equally to this work.}
	\affiliation{Institute of Materials Science, Technical University of Darmstadt, Darmstadt 64287, Germany}
	
	\author{Vikrant Chaudhary}
    \affiliation{Institute of Materials Science, Technical University of Darmstadt, Darmstadt 64287, Germany}
	\affiliation{Physics Department and CSMB, Humboldt-Universität zu Berlin, 12489 Berlin, Germany}
	
	\author{Shengqiao Wang}
	\affiliation{School of Materials Science and Engineering, Jilin University, Changchun 130012, China}
	
	\author{Chen Shen}
    \email{shenc@szlab.ac.cn}
	\affiliation{Suzhou Laboratory, Suzhou, 215021, China} 

	\author{\\Hao Wang}
	\email{haowang@tmm.tu-darmstadt.de}
	\affiliation{Institute of Materials Science, Technical University of Darmstadt, Darmstadt 64287, Germany}
	
	\author{Hongbin Zhang}
	\affiliation{Institute of Materials Science, Technical University of Darmstadt, Darmstadt 64287, Germany}
	
	\begin{abstract}
		Altermagnets provide a promising platform for a wide spectrum of applications integrating advantages of conventional ferromagnets and antiferromagnets. In this work, we implement a medium-throughput first-principles workflow and evaluate topological transport properties driven by quantum geometry for 135 altermagnets in the MAGNDATA database. Based on automated Wannier construction, both linear and nonlinear responses, including the anomalous Hall effect, magneto-optical Kerr effect, and bulk photovoltaic effect, are evaluated with further symmetry verifications. Detailed analysis is done on representative cases like metallic \ce{VNb3S6} with enhanced anomalous Hall conductivity, \ce{CaIrO3} with giant MOKE, and \ce{CuFeS2} with large shift current in non-centrosymmetric. These results establish a symmetry-guided computational route for identifying experimentally accessible fingerprints and functional transport properties in altermagnets.
	\end{abstract}
	
	\maketitle

\section{Introduction}

Altermagnetism has recently emerged as a new frontier beyond conventional ferromagnetism and antiferromagnetism in condensed matter physics, and altermagnets (ALTs) exhibit fascinating properties \cite{vsmejkal2022beyond}. Despite vanishing net magnetization, ALTs host pronounced momentum-dependent non-relativistic spin splittings (NRSSs) even in the absence of spin–orbit coupling (SOC) \cite{vsmejkal2022emerging}. Such unusual coexistence of compensated magnetism and spin polarization renders ALTs attractive for spintronic and optoelectronic applications because of their negligible stray fields, robustness against magnetic cross-talk, and electronically active spin splitting \cite{bai2024altermagnetism}. 

Spin space group theory provides a comprehensive description of altermagnetic spin splitting \cite{jungwirth2026symmetry}, the existence of possible physical responses, as well as tools to identify possible ALT candidates with given magnetic configurations. However, quantitative assessments on NRSS and associated properties are indispensable to screen for materials with potential applications. 
In this regard, high-throughput (HTP) calculations are essential. 
For instance, advanced workflows combining density functional theory (DFT) and dynamical mean field theory (DMFT) have uncovered previously unknown metallic ALTs \cite{wan2025high}, and HTP DFT calculations have been performed to evaluate the magnitudes of NRSSs \cite{guo2023spin,chen2025unconventional,sufyan2025high}. All these works are valuable to guide further experimental validation of the ALT nature of corresponding materials. Nevertheless, to bridge to possible applications, the transport properties of ALTs, in particular the topological transports in both linear and nonlinear regimes, remain largely unexplored.

Closing this gap requires a framework that treats symmetry, electronic structure, and transport on an equal footing. In this regard, intrinsic transport phenomena caused by the quantum geometry of Bloch states, including the Berry curvature and quantum metric, are strongly regulated by underlying crystal structures and symmetries \cite{verma2026quantum, liu2023covariant, qiang2026quantum}. For instance, in ALTs, this geometric perspective is particularly compelling: the absence or breaking of combined symmetries such as $\mathcal{PT}$, together with the transformation properties of time-reversal and spin-space operations, can strongly constrain the allowed Berry curvature, quantum metric, and their higher-order moments \cite{vsmejkal2022anomalous,wang2023quantum}. ALTs therefore provide an ideal setting in which to explore how symmetry and magnetic order reshape the phase texture of Bloch wave functions and its encoding in experimentally observable transport and optical responses.
This connection is most directly reflected in the linear-response regime, where phenomena such as the anomalous Hall effect (AHE) \cite{gonzalez2023spontaneous,zhou2025manipulation,wang2025symmetry}, anomalous Nernst effect (ANE) \cite{li2025large,zhou2024crystal}, spin Hall effect 
\cite{liao2024separation,jeong2026magnetic} magneto-optical Kerr effect (MOKE) \cite{pan2026experimental,luo2026symmetry}, and x-ray magnetic circular dichroism (XMCD) \cite{xie2025x,hariki2024x} arise from Berry curvature. Beyond linear response, ALTs also host a rich variety of nonlinear transport effects, including the shift current \cite{gu2025ferroelectric,sivianes2025optical,yang2025giant}, injection current \cite{jiang2025nonlinear,dong2025crystal}, second harmonic generation \cite{ma2025probing}, and nonlinear Hall effect \cite{zhu2025magnetic}, which are linked to higher-order geometric quantities such as the Berry-curvature dipole and the quantum metric. Therefore, ALTs are hosting not merely a new magnetic phase but also a symmetry-engineered platform for realizing and controlling a broad spectrum of topological, optical, and spintronic functionalities.

In this work, we perform medium-throughput DFT calculations to evaluate the topological transport properties driven by quantum geometry in ALTs. Starting from a symmetry-curated set of altermagnetic compounds,  automated Wannier interpolation and symmetry analysis are integrated with DFT to evaluate linear and nonlinear intrinsic topological responses, including the AHE, ANE, MOKE, and bulk photovoltaic effect (BPVE). Beyond identifying candidate materials with pronounced properties, we aim to uncover how altermagnetic order, quantum geometry, and symmetry jointly regulate such experimentally accessible fingerprints. By analyzing these complementary responses, it is demonstrated that transport and optical observables offer a practical route to characterize altermagnetism in real materials. Our framework provides symmetry-guided design principles for the discovery and engineering of high-performance altermagnetic materials.

\begin{figure}
	\centering
	\includegraphics[width=0.98\linewidth]{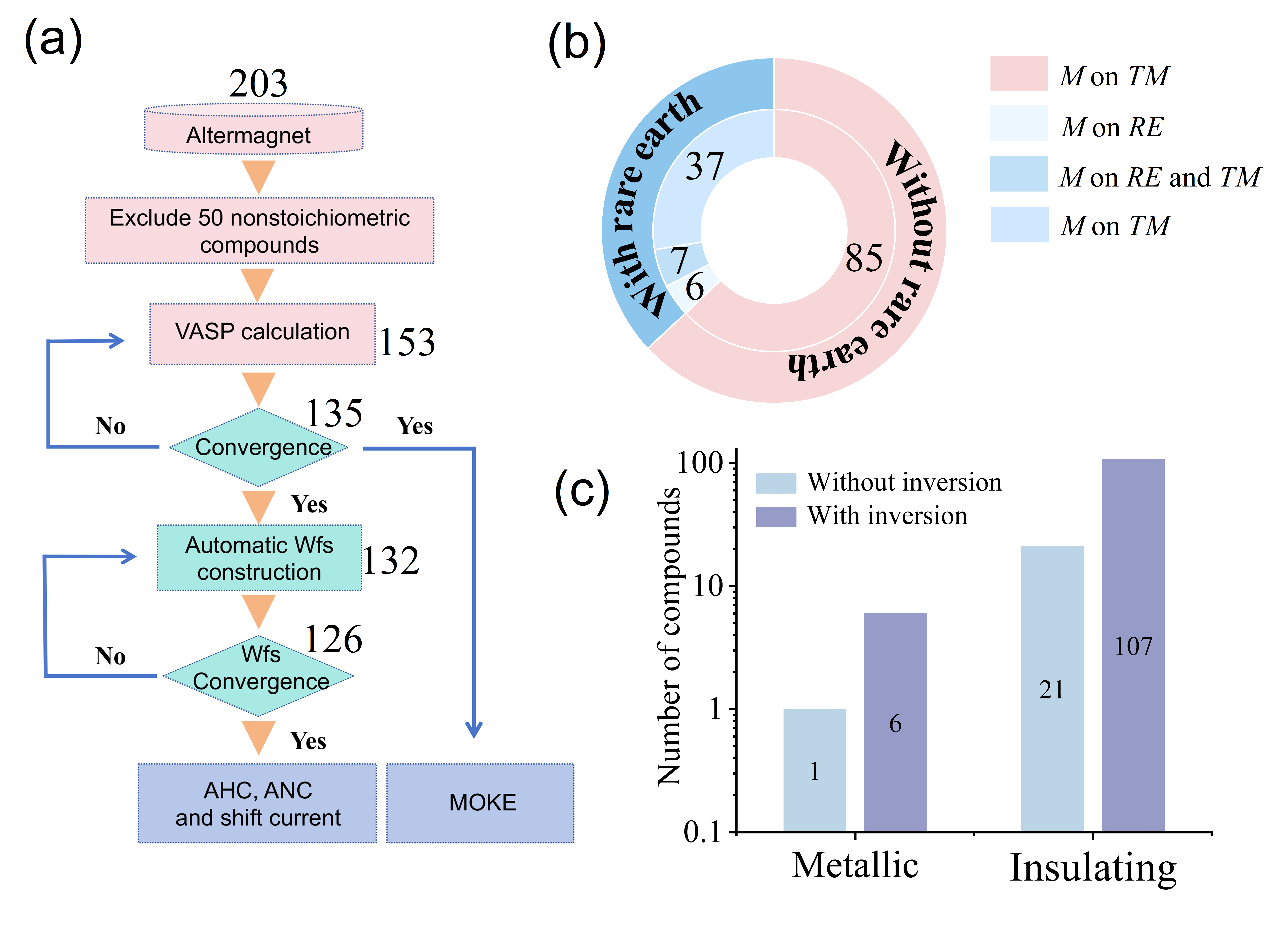}
	\caption{\label{fig:workflow}Workflow of the medium-throughput screening of altermagnetic materials. (a) Flowchart illustrating the step-by-step filtering and computational procedure, with the number of remaining compounds indicated at each stage. (b) Distribution of magnetic elements ($M$) in the screened compounds, divided into systems without rare-earth elements (RE) and those containing RE. The RE-containing compounds are further classified according to whether the magnetic atoms are located on RE sites, transition-metal (TM) sites, or both. (c) Classification of the final set of compounds according to their electronic character (metallic or insulating) and the presence or absence of inversion symmetry.}
\end{figure}

\section{Results and Discussion}

\subsection{Workflow}

To systematically evaluate transport and optical responses in altermagnetic materials, we constructed a screening workflow based on first-principles calculations and automated Wannier interpolation, as illustrated in Figure~\ref{fig:workflow}. Our starting point is the MAGNDATA database \cite{gallego2016magndata}, which provides experimentally reported crystal structures and magnetic configurations. From this database, we collected 203 altermagnetic compounds previously identified by DFT calculations and classified by spin space group \cite{chen2025unconventional,zhu2025magnetic}. After excluding 50 non-stoichiometric entries, 153 stoichiometric compounds were retained for further study.

All candidate materials were first examined by DFT+$U$ calculations including SOC. Hubbard corrections were applied to localized transition-metal $d$ states when relevant \cite{moore2024high}. For rare-earth-containing compounds, the $4f$ electrons were either treated within the $4f$-in-core approximation when magnetism is dominated by transition-metal atoms, or explicitly included in the valence with an effective Hubbard $U_{\mathrm{eff}}$ when rare-earth moments participate in the magnetic order \cite{jiang2009localized}. The specific Hubbard parameters used for each compound are listed in Table~S2. After convergence tests on total energies, atomic forces, and magnetic moments, 18 compounds were discarded, leaving 135 converged candidates.
We then carried out automated maximally localized Wannier function (MLWF) construction using an in-house wannierization workflow interfacing VASP and Wannier90 \cite{zhang2018high}. Among the 135 converged systems, three were excluded at this stage because the memory requirements exceeded the available computational resources, and 132 compounds eventually yielded well-converged Wannier representations. These Wannier Hamiltonians provide the common basis for evaluating the response functions discussed below. Metallic systems were analyzed in terms of Berry-curvature-driven linear responses, including the AHE and ANE. Insulating candidates were further examined for the MOKE, while non-centrosymmetric compounds were selected for second-order shift-current calculations. In this way, the workflow links the symmetry classification of ALTs to the corresponding linear and nonlinear transport observables and provides the basis for the material-specific analyses presented in the following sections.

\begin{table*}[htbp]
	\centering
	\caption{AHE in representative metallic altermagnet candidates.
For each compound, the table lists the chemical formula, Néel-vector orientation, magnetic space group, symmetry-allowed AHC tensor components, and the corresponding reference.}
	\label{tab:AHC}
	
	\begin{tabular}{c c c c c c}
		\toprule
		ID & Chemical formula & Néel vector & Magnetic space group  
		& AHC tensor & Reference \\
		\midrule
		
		0.118 & \ce{Ba5Co5ClO13} & $[001]$ & $194.268\,(P6'_3/m'm'c')$ & $(0, 0, 0)$ & Ref.~\cite{yamaura2001synthesis} \\
		0.358 & \ce{CaFe5O7} & $[010]$ & $11.54\,(P2'_1/m')$ & $(\sigma_{yz}, 0, \sigma_{xy})$ & Ref.~\cite{wang2025large} \\
		0.528 & \ce{CrSb} & $[001]$ & $194.268\,(P6'_3/m'm'c')$ & $(0, 0, 0)$ & Ref.~\cite{zhou2025manipulation} \\
		0.607 & \ce{RuO2} & $[001]$ & $136.499\,(P4'2/mnm')$ & $(0, 0, 0)$ & Ref.~\cite{fedchenko2024observation} \\
		0.708 & \ce{CrNb4S8} & $[001]$ & $194.268\,(P6'_3/m'm'c')$ & $(0, 0, 0)$ & Ref.~\cite{chen2025unconventional} \\
		0.712 & \ce{VNb3S6} & $[100]$ & $20.33\,(C2'2'2_1)$ & $(0, 0, \sigma_{xy})$ & Ref.~\cite{zhu2025magnetic} \\
		0.999 & \ce{Fe4O5} & $[001]$ & $36.174\,(Cm'c2'_1)$ & $(0, \sigma_{xz}, 0)$ & Ref.~\cite{yang2021metallic} \\
		\bottomrule
	\end{tabular}
\end{table*}

\subsection{Metallic ALTs}

For ALTs with collinear magnetic ordering, emergence of a spontaneous AHE can be attributed to SOC, thus is subject to the magnetic space group. Therefore, the Néel-vector orientation is critical \cite{vsmejkal2022anomalous}.
Table~\ref{tab:AHC} summarizes the symmetry-allowed anomalous Hall conductivity (AHC) components in metallic ALTs under different magnetic space groups. Obviously, only a subset of the metallic ALTs exhibit finite AHC under the experimentally relevant magnetic configuration, whereas the others are symmetry-forbidden and therefore exhibit a vanishing spontaneous Hall response. In particular, \ce{CaFe5O7} with the Néel vector along [010] and magnetic space group $P2_1'/m'$ allows two off-diagonal components, $(\sigma_{yz},\sigma_{xy})$; \ce{VNb3S6} with magnetic space group $C2'2'2_1$ allows a finite $\sigma_{xy}$ component; and \ce{Fe4O5} with magnetic space group $Cm'c2_1'$ allows a finite $\sigma_{xz}$ component. 
By contrast, \ce{Ba5Co5ClO13}, \ce{CrSb}, \ce{RuO2}, and \ce{CrNb4S8} belong to magnetic space groups whose symmetry operations enforce complete cancellation of the anomalous Hall tensor, resulting in zero AHC. Importantly, these symmetry constraints are not immutable: a spontaneous AHC can be induced by rotating the Néel vector from the out-of-plane direction to $\tfrac{1}{2}(a+b)$ in \ce{CrSb} \cite{yu2025neel}, or by applying uniaxial strain to lower the symmetry in \ce{RuO2} \cite{liang2025strain}. Such perturbations activate the anomalous Hall response by modifying the underlying magnetic symmetry, highlighting the strong sensitivity of the AHE in ALTs to Néel-vector orientation. Consequently, magnetic space group analysis serves as a powerful tool for predicting topological transport properties and for efficiently identifying altermagnetic materials with spontaneous anomalous Hall responses.

\begin{figure*}
	\centering
	\includegraphics[width=0.9\linewidth]{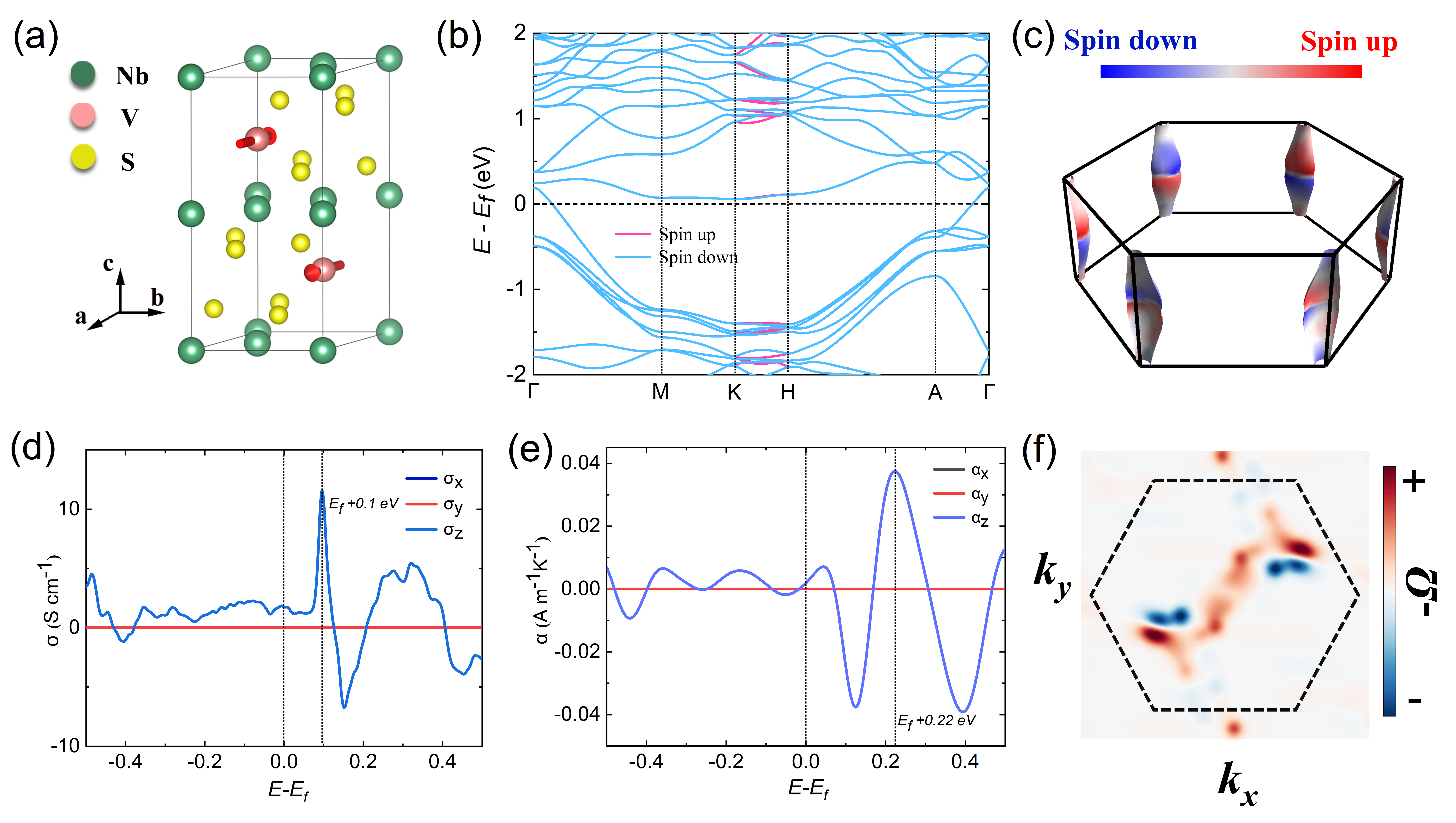}
	\caption{\label{fig:AHC}
		(a) Crystal structure of the altermagnetic compound \ce{VNb3S6}. Green, pink, and yellow spheres denote Nb, V, and S atoms, respectively.
(b) Spin-resolved electronic band structure along the high-symmetry path of the Brillouin zone, where the spin-up and spin-down characters are shown in pink and blue, respectively.
(c) Spin-resolved constant-energy-surface cross section taken at 1.4 eV below the Fermi level, revealing the alternating spin-polarization pattern along the K–H direction. 
(d) Energy-dependent AHC as a function of the Fermi level.
(e) Energy-dependent ANC as a function of the Fermi level.
(f) $k$-resolved Berry-curvature distribution $-\Omega_{z}$ in the $k_z=0$ plane at 0.1 eV above the Fermi level.}
\end{figure*}

To further elucidate the microscopic origin of the spontaneous AHE, we selected \ce{VNb3S6} as a representative example, as shown in Figure~\ref{fig:AHC}. Figure~\ref{fig:AHC}(a) presents its three-dimensional crystal structure, in which the magnetic moments are parallel within each V layer and antiparallel between neighboring V layers, while the Néel vector is oriented along the $a$ axis \cite{lu2020canted,zhu2025magnetic}. This collinear antiferromagnetic configuration belongs to the $g$-wave altermagnetic class. In the SOC-free limit, the corresponding spin-space group is $P^{-1}6_3^{-1}2_1 2^{\infty m}1$, whose spin-space operations connect the opposite-spin sublattices and impose a characteristic alternating spin polarization in momentum space. The spin-only subgroup associated with continuous spin rotations around the Néel vector further forbids a pure charge AHE without SOC. The corresponding band structure without SOC is shown in Figure~\ref{fig:AHC}(b). Despite the vanishing net magnetization, pronounced momentum-dependent spin splitting appears in the electronic bands, which is a defining feature of altermagnetism. Figure~\ref{fig:AHC}(c) further shows that the spin-resolved constant energy surface cross-section of \ce{VNb3S6} at $1.4~\mathrm{eV}$ below the Fermi level exhibits an alternating spin-polarization pattern along the K--H direction, consistent with the symmetry-imposed $g$-wave spin splitting. Guided by the symmetry analysis, we investigate the energy dependence of the nonzero AHC and anomalous Nernst conductivity (ANC) components, $\sigma_{xy}$ and $\alpha_{xy}$, as shown in Figure~\ref{fig:AHC}(d) and (e), respectively. Both quantities exhibit sharp peaks near the Fermi level, reaching maximum values of approximately $10~\mathrm{S/cm}$ for the AHC and $0.04~\mathrm{A m^{-1}K^{-1}}$ for the ANC. These results indicate that doping and strain can significantly enhance the anomalous Hall response \cite{shen2025multifunctionality,smolyanyuk2025origin,furuhashi2025doping}. To clarify the origin of the Hall peak, we plot the $k$-resolved Berry-curvature distribution $-\Omega_z$ in the Brillouin zone at $E_{\mathrm F}+0.1~\mathrm{eV}$, as shown in Figure~\ref{fig:AHC}(f). Most regions of the two-dimensional Brillouin zone contribute positively, consistent with the sign of the integrated AHC, while the coexistence of positive and negative regions reflects the symmetry-imposed Berry-curvature texture of the altermagnetic state. Since the pure charge AHC is forbidden in the SOC-free spin-space-group limit, the finite $\sigma_{xy}$ obtained here directly demonstrates that SOC converts the underlying altermagnetic band splitting into an observable Hall signal by generating a nonvanishing Berry-curvature distribution. These results establish \ce{VNb3S6} as a representative metallic altermagnet with pronounced transverse transport responses and illustrate how SOC and symmetry together govern the emergence of Hall and thermoelectric effects in metallic ALTs.

\subsection{Insulating ALTs}

In contrast to the metallic systems discussed above, where the AHC can be directly evaluated at the Fermi level, insulating compounds require a different approach to probe transverse responses. In this case, the MOKE can be understood as a manifestation of the frequency-dependent AHC. According to Equation~\ref{eq:opt_conductivity}, for an insulator, the dc limit $\omega \to 0$ of the off-diagonal component $\sigma_{xy}(\omega)$ vanishes in the absence of a quantized Hall response, but its finite-frequency value is controlled entirely by interband optical transitions between occupied and unoccupied states. The Kerr rotation and ellipticity are then directly related to this ac Hall conductivity via Equation~\ref{eq:moke_sigma}. The non-vanishing components are dictated by magnetic point-group symmetry, while its magnitude and spectral structure arise from the interband optical matrix elements in the expression above.

\begin{figure}
\centering
\includegraphics[width=0.9\linewidth]{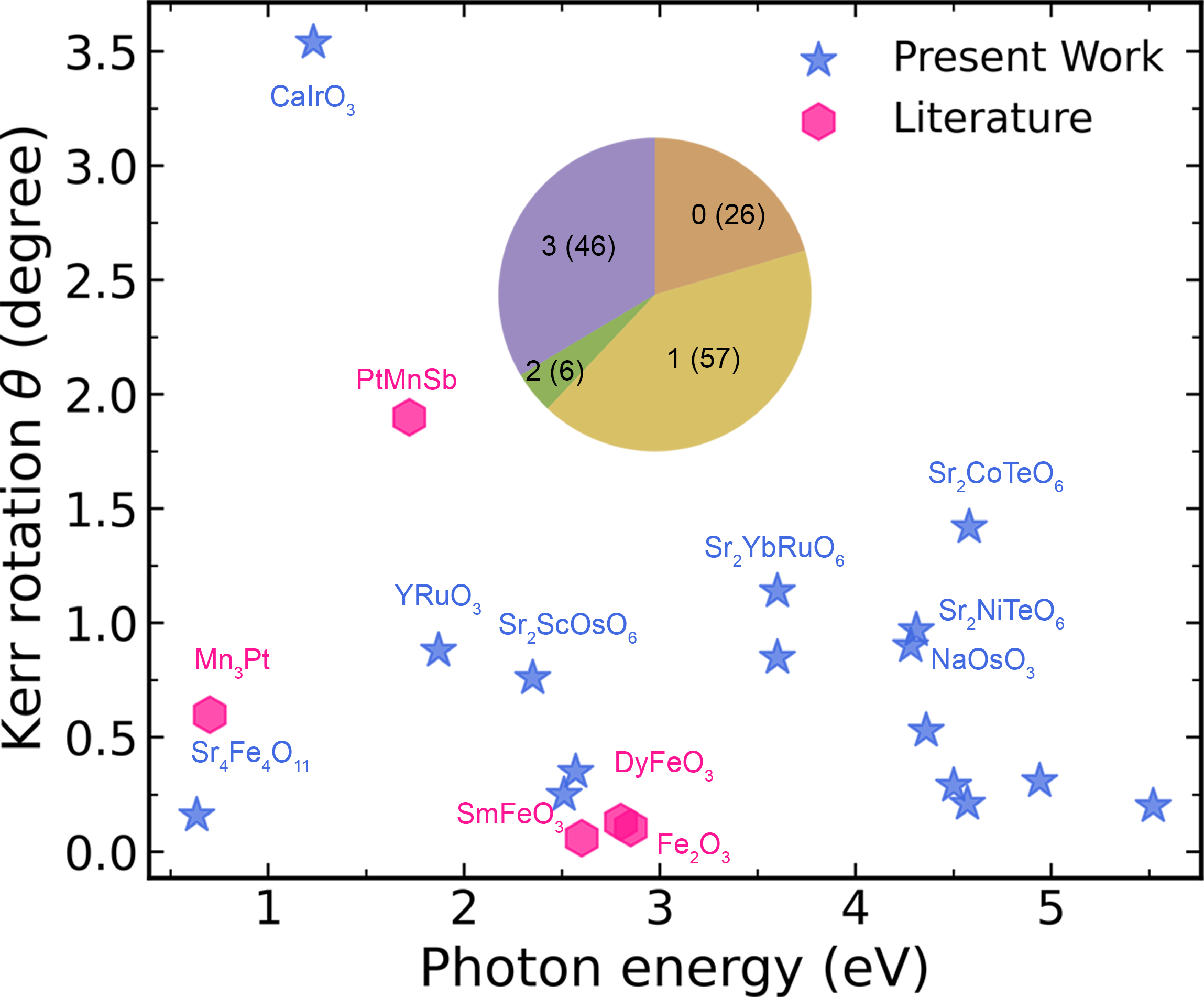}
\caption{Kerr rotation angle as a function of photon energy. Pie chart showing the distribution of the number of MOKE tensor elements. Blue stars denote the results from the present work, while purple diamonds represent values reported in the literature.}
\label{fig:moke}
\end{figure}

It therefore provides a powerful route for accessing transverse responses in insulating systems. The symmetry-allowed Kerr tensor components are fixed by the magnetic space group, while the magnitude of the Kerr rotation shows a strong material dependence. We further compiled representative insulating ALTs with sizable MOKE responses, as listed in Table~\ref{tab:MOKE}. A general trend emerging from our dataset is that the magnitude of the MOKE response correlates strongly with the presence of heavy elements and thus the overall strength of SOC. As observed in Figure~\ref{fig:moke}, large MOKE responses are mainly found in compounds containing heavier $4d$ and $5d$ elements, such as \ce{NaOsO3}, \ce{YRuO3}, \ce{Sr2YbRuO6}, \ce{Sr2ScOsO6}, \ce{Sr2NiTeO6}, and \ce{Sr2CoTeO6}. However, this trend does not show a simple one-to-one correspondence with the size of the local orbital magnetization~\cite{koMagnetoopticalObservationElectrically2026}.
For instance, both \ce{Sr2YbRuO6} and \ce{Sr2CoTeO6} exhibit large Kerr rotations exceeding $1^\circ$, despite a significant difference in their orbital moments. In \ce{Sr2YbRuO6}, the Ru site carries only a small orbital moment of about $0.02\ \mu_B$, whereas in \ce{Sr2CoTeO6}, the Co site exhibits a much larger orbital moment of approximately $0.2\ \mu_B$. As shown in Figure~S11, the two systems exhibit distinct microscopic origins of the Kerr response. \ce{Sr2YbRuO6} shows more pronounced altermagnetic spin splitting (Figure~S11(a)), leading to a robust Kerr signal that is largely insensitive to SOC strength, as evidenced by the nearly identical $\theta_{zx}(\omega)$ spectra for $\lambda=1.0$ and $0.3$ (Figure~S11(c)). In contrast, \ce{Sr2CoTeO6} displays weaker spin splitting (Figure~S11(b)) but a much stronger dependence of $\theta_{zx}(\omega)$ on SOC (Figure~S11(d)), indicating a SOC-dominated mechanism consistent with its larger orbital moment. These results demonstrate that large Kerr rotation can originate either from SOC-enhanced orbital polarization or from altermagnetic band reconstruction in weak-orbital-moment systems.

\begin{table*}[htbp]
\centering
\caption{MOKE in representative insulating altermagnet candidates under their specified magnetic configurations. For each compound, the table summarizes the chemical formula, Néel-vector orientation, magnetic space group, symmetry-allowed MOKE tensor components, and the corresponding maximum Kerr rotation angles $\theta_{\max}$.
}
\label{tab:MOKE}

\begin{tabular}{c c c c c c}
\toprule
ID & Chemical formula & Néel vector & Magnetic space group  
& MOKE tensor & $\theta_{\mathrm{max}}$ (deg) \\
\midrule

0.128 & \ce{FeSO4F} & $[010]$ & $15.89\,(C2'/c')$ & ($\theta_{yz}$, 0, $\theta_{xy}$) & (0.43, 0, 0.52) \\

0.25 & \ce{NaOsO3} & $[001]$ & $62.448\,(Pn'ma')$ & (0, $\theta_{zx}$, 0) & (0, 0.90, 0) \\

0.379 & \ce{SmFeO3} & $[001]$ & $62.446\,(Pn'm'a)$ & ($\theta_{yz}$, 0, 0) & (0.25, 0, 0) \\

0.380 & \ce{SmFeO3} & $[100]$ & $62.448\,(Pn'ma')$ & (0, 0, $\theta_{xy}$) & (0, 0, 0.21) \\

0.402 & \ce{Sr4Fe4O11} & $[010]$ & $65.486\,(Cmm'm')$ & ($\theta_{yz}$, 0, 0) & (0.16, 0, 0) \\

0.513 & \ce{YRuO3} & $[001]$ & $62.448\,(Pn'ma')$ & (0, $\theta_{zx}$, 0) & (0, 0.88, 0) \\

0.669 & \ce{Sr2YbRuO6} & $\begin{array}{c}
[010]+\\
39^\circ [100]
\end{array}$ &  $14.75\,(P2_1/c)$ & (0, 0, $\theta_{xy}$) & (0, 0, 0.85) \\

0.670 & \ce{Sr2YbRuO6} & $\begin{array}{c}
[001]+\\
23^\circ [100]
\end{array}$ & $14.75\,(P2_1/c)$ & (0, $\theta_{zx}$, 0) &(0, 1.14, 0) \\

0.65 & $\alpha$-\ce{Fe2O3} & $[100]$ & $15.89\,(C2'/c')$ & (0, $\theta_{zx}$, $\theta_{xy}$) & (0, 0.20, 0.02) \\

0.755 & \ce{Mn2SeO3F2} & $[001]$ & $62.448\,(Pn'ma')$ & (0, $\theta_{zx}$, 0) & (0, 0.29, 0) \\

0.760 & \ce{FeOHSO4} & $[010]$ & $15.89\,(C2'/c')$ & ($\theta_{yz}$, 0, $\theta_{xy}$) & (0.31, 0, 0.32) \\

0.79 & \ce{CaIrO3} & $[001]$ & $63.464\,(Cm'cm')$ & (0, $\theta_{zx}$, 0) & (0, 3.54, 0) \\

0.836 & \ce{DyFeO3} & $[100]$ & $62.448\,(Pn'ma')$ & (0, 0, $\theta_{xy}$) & (0, 0, 0.35) \\

0.917 & \ce{Sr2ScOsO6} & $[100]$ & $14.75\,(P2_1/c)$ & (0, $\theta_{zx}$, 0) & (0, 0.76, 0) \\

0.934 & \ce{Sr2NiTeO6} & $\begin{array}{c}
[001]+\\
41^\circ [100]
\end{array}$ & $14.75\,(P2_1/c)$ & (0, $\theta_{zx}$, 0) & (0, 0.97, 0) \\

0.937 & \ce{Sr2CoTeO6} & $\begin{array}{c}
[001]+\\
32^\circ [100]
\end{array}$ & $14.75\,(P2_1/c)$ & (0, $\theta_{zx}$, 0) & (0, 1.42, 0) \\

\bottomrule
\end{tabular}
\end{table*}

\begin{figure*}
\centering
\includegraphics[width=0.90\linewidth]{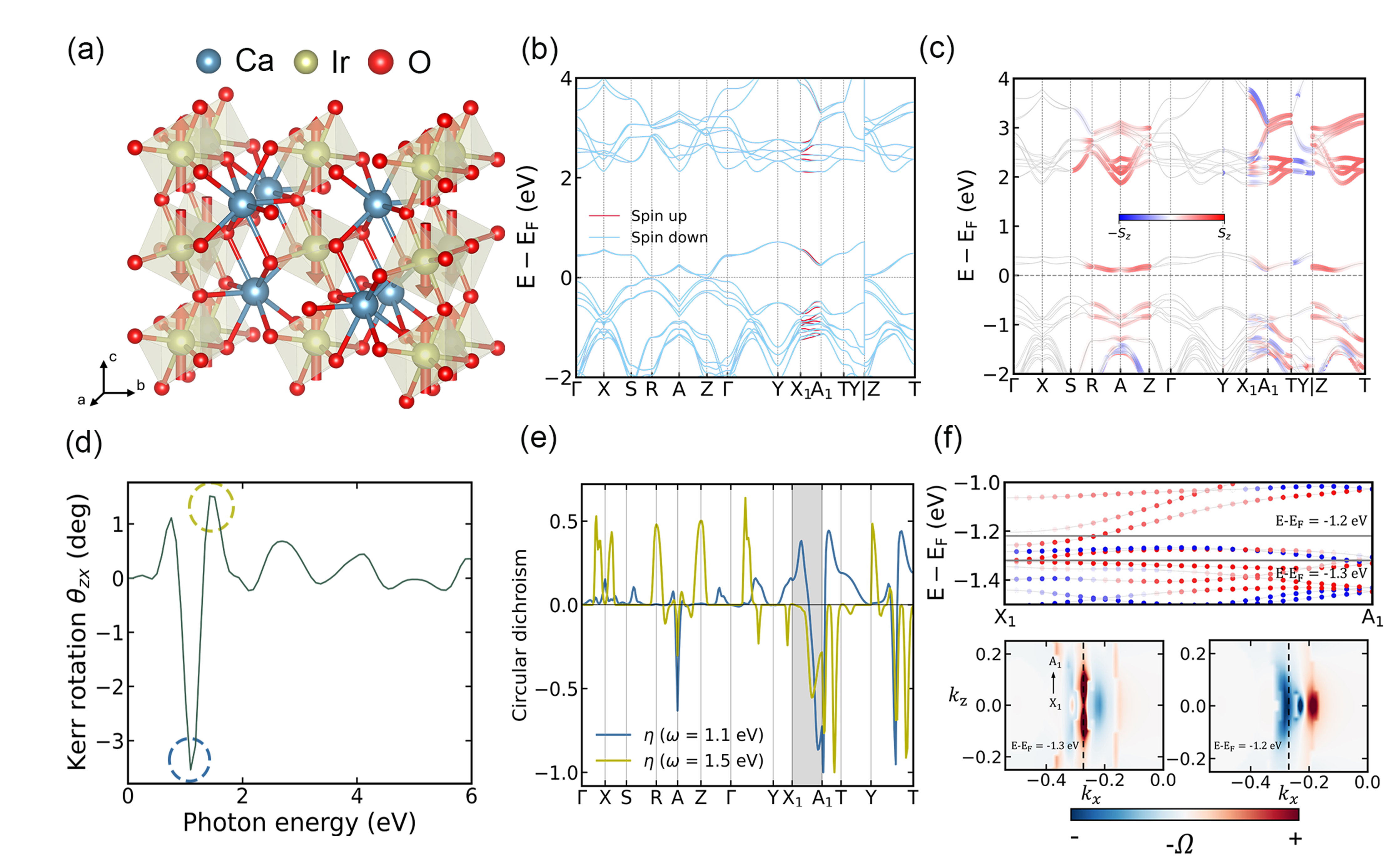}
\caption{(a) Crystal structure of the insulating altermagnetic compound \ce{CaIrO3} with magnetic space group $Cm'cm'$.
(b) Spin-polarized band structure of \ce{CaIrO3}.
(c) Band structure of \ce{CaIrO3} with SOC under the aforementioned magnetic space group.
(d) Kerr rotation angle $\theta_{zx}$ as a function of photon energy.
(e) $k$-path resolved circular dichroism $\eta$ at the Kerr peaks highlighted in (d). Shaded area indicates the path where NRSS occurs.
(f) (top) Band structure with SOC along $X_1 -A_1$ where the altermagnetic spin splitting is allowed. (bottom) Heat map of the Berry curvature on the $k_x$-$k_z$ plane ($k_y$ = 0.72) at Fermi energy $E - E_F = -1.3$ eV (left) and $E - E_F = -1.2$ eV (right), respectively. Dashed lines indicate the $k$ path along $X_1 -A_1$.
}
\label{fig:CAIRO3}
\end{figure*}

Among all insulating candidates, \ce{CaIrO3} stands out by exhibiting the largest Kerr rotation in our dataset, with a symmetry-allowed $\theta_{zx}$ reaching up to $3.5^\circ$. This unusually large response originates from the interplay of strong SOC, crystal-field effects, and electronic correlations in the Ir $5d$ manifold. SOC entangles the $t_{2g}$ states and, together with the Hubbard interaction, stabilizes a spin–orbit-assisted $j_{\mathrm{eff}}=1/2$ insulating state~\cite{kimNovel$J_mathrmeff12$2008,sala$mathrmCaIrO_3$SpinOrbitMott2014,zhangEffective$Jmathbf12$2013}. Our DFT+U calculations yield a band gap of about 0.5 eV. Despite this moderate gap, optical transitions from hybridized $t_{2g}$ valence states to conduction states near the band edge produce strong Kerr signals in the visible range, facilitated by substantial interband matrix elements. In realistic structures, octahedral tilting and rotation further mix the $t_{2g}$ orbitals, driving the system away from the idealized limit and enhancing optical activity. As shown in Figure~\ref{fig:CAIRO3}(a), \ce{CaIrO3} crystallizes in a post-perovskite structure of corner-sharing \ce{IrO6} octahedra with alternating tilts of approximately $\pm 23^\circ$, forming zigzag Ir–O–Ir chains and inducing a small canting of the magnetic moments toward the $b$ axis, consistent with experiment \cite{ohgushiResonantXrayDiffraction2013b}. This distortion breaks the combined $\mathcal{PT}$ symmetry and connects magnetic sublattices via a $\{ C_{2y} | 00\frac{1}{2} \}$ operation, thereby enabling symmetry-allowed altermagnetic spin splitting shown in Figure.~\ref{fig:CAIRO3}(b). First-principles calculations show that a momentum-dependent splitting already appears in the non-relativistic band structure along the $X_1 \rightarrow A_1$ path, in agreement with spin-space-group analysis. Including SOC (Figure~\ref{fig:CAIRO3}(c)) opens the insulating gap and lifts remaining degeneracies, while preserving the characteristic altermagnetic splitting. Notably, the splitting develops primarily along $k_z$, whereas SOC-induced spin polarization lies mainly in the $k_x$–$k_y$ plane, indicating a separation between altermagnetic and SOC-driven spin textures in both momentum and spin space.

The altermagnetic electronic structure gives rise to a pronounced MOKE response $\theta_{zx}$, reaching $3.5^\circ$ around 1.1 eV and $1.26^\circ$ around 1.5 eV, as shown in Figure~\ref{fig:CAIRO3}(d). These peaks originate from interband transitions between valence and conduction states with distinct orbital character, leading to strong circular dichroism. This is confirmed in Figure~\ref{fig:CAIRO3}(e), where the $k$-path-resolved circular dichroism $\eta$ is shown for photon energies corresponding to the two Kerr peaks. In both cases, large dichroic signals are concentrated along the shaded region of the $k$ path, where the altermagnetic spin splitting is allowed by symmetry.
To further elucidate the role of this splitting in the MOKE response, we compute the two-dimensional Berry curvature on a plane containing the splitting path. As shown in Figure~\ref{fig:CAIRO3}(f), the band structure along the $X_1 \rightarrow A_1$ path (upper panel) exhibits a clear altermagnetic spin splitting that persists in the presence of SOC. The corresponding Berry curvature maps at two representative energies (lower panels), $E - E_F = -1.3$ eV and $-1.2$ eV, demonstrate that this high-symmetry path (marked by the vertical dashed line) coincides with regions of pronounced Berry curvature. In particular, near the energy window of the altermagnetic splitting ($E - E_F = -1.3$ eV), large Berry curvature with opposite signs develops on the two sides of the splitting. At a slightly higher energy ($E - E_F = -1.2$ eV), a similarly strong but redistributed Berry curvature pattern is observed.
These results establish a direct connection between the $k$-resolved circular dichroism and the Berry curvature distribution: the altermagnetic band splitting enhances the asymmetry of interband optical transitions along specific $k$-space paths, which in turn generates substantial Berry curvature. Consequently, beyond the conventional SOC-driven mechanism, the altermagnetic splitting itself provides an additional and efficient contribution to the MOKE response.

\subsection{Inversion-asymmetric ALTS}

\begin{table*}[htbp]
\centering
\caption{BPVE in inversion-asymmetric altermagnet candidate materials.
For each compound, the table lists the chemical formula, Néel-vector orientation, crystallographic point group, band gap, maximal shift-current response, and symmetry-allowed shift-current tensor components.}
\label{tab:shiftcurrent}

\begin{tabular}{c c c c c c c c}
\toprule
ID & \makecell {Chemical \\ formula} & \makecell {Néel \\ vector} & \makecell {Point \\ group}
& \makecell{Band gap \\(eV) }
& \makecell{Maximal \\ Shift current \\($\mu$A/V$^2$)} & \makecell {Symmetry allowed \\ shift current components} \\
\midrule

0.1018 & \ce{SrMnO3} & $[100]$ & $222$ & 1.81 & 8.55 & $\sigma^{xyz},\ \sigma^{yxz},\ \sigma^{zxy}$\\
0.1019 & \ce{SrMnO$_3$} & $[010]$ & $222$  & 1.81 & 8.55 & $\sigma^{xyz},\ \sigma^{yxz},\ \sigma^{zxy}$\\
0.137 & \ce{Cu$_2$V$_2$O$_7$} & $[100]$ & $mm2$  & 2.08 & 6.8 &$\sigma^{zxx},\ \sigma^{zyy},\ \sigma^{zzz},\ \sigma^{yyz},\ \sigma^{xxz}$\\
0.21 & \ce{PbNiO$_3$} & $[001]$ & $3m$  & 0.60 & 12.5 & \makecell[l]{
$\sigma^{xxz}=\sigma^{yyz}$, $\sigma^{zxx}=\sigma^{zyy}$, \\ $\sigma^{xxy}=\sigma^{yxx}=-\sigma^{yyy}$, $\sigma^{zzz}$
}\\
0.229 & \ce{Ba$_2$MnSi$_2$O$_7$} & $[001]$ & $\bar42m$  & 3.34 & -3.4 & $\sigma^{xyz}=\sigma^{yxz},\ \sigma^{zxy}$\\
0.23 & \ce{Ca$_3$Mn$_2$O$_7$} & $[100]$ & $mm2$ & 1.14 & -19.65 & $\sigma^{zxx},\ \sigma^{zyy},\ \sigma^{zzz},\ \sigma^{yyz},\ \sigma^{xxz}$\\
0.241 & \ce{Y$_2$Cu$_2$O$_5$} & $[010]$ & $ mm2$  & 1.95 & -4.26 & $\sigma^{zxx},\ \sigma^{zyy},\ \sigma^{zzz},\ \sigma^{yyz},\ \sigma^{xxz}$\\
0.254 & \ce{[C(ND$_2$)$_3$]Cu(DCOO)$_3$} & $[010]$ & $mm2$ & 2.47 & 2.98 & $\sigma^{zxx},\ \sigma^{zyy},\ \sigma^{zzz},\ \sigma^{yyz},\ \sigma^{xxz}$\\
0.255 & \ce{[C(ND$_2$)$_3$]Cu(DCOO)$_3$} & $[100]$ & $mm2$& 2.47 & 2.97 &$\sigma^{zxx},\ \sigma^{zyy},\ \sigma^{zzz},\ \sigma^{yyz},\ \sigma^{xxz}$\\
0.303 & \ce{BaCrF$_5$} & $[001]$ & $222$ &  3.72 & -0.05 & $\sigma^{xyz},\ \sigma^{yxz},\ \sigma^{zxy}$\\
0.331 & \ce{Fe$_2$Mo$_3$O$_8$} & $[001]$ & $6mm$  & 0.77 & 23.2 & $\sigma^{xxz}=\sigma^{yyz},\ \sigma^{zxx}=\sigma^{zyy},\ \sigma^{zzz}$\\
0.332 & \ce{Co$_2$Mo$_3$O$_8$} & $[001]$ & $6mm$ & 1.63 & -17.2 &$\sigma^{xxz}=\sigma^{yyz},\ \sigma^{zxx}=\sigma^{zyy},\ \sigma^{zzz}$\\
0.338 & \ce{Co$_2$Mo$_3$O$_8$} & $[001]$ & $6mm$ & 1.70 & 13.9 &$\sigma^{xxz}=\sigma^{yyz},\ \sigma^{zxx}=\sigma^{zyy},\ \sigma^{zzz}$\\
0.50 & \ce{MnTiO$_3$} & $[100]$ & $3m$ & 2.59 & -11.2 & \makecell[l]{
$\sigma^{xxz}=\sigma^{yyz}$, $\sigma^{zxx}=\sigma^{zyy}$, \\ $\sigma^{xxy}=\sigma^{yxx}=-\sigma^{yyy}$, $\sigma^{zzz}$
}\\
0.56 & \ce{Ba$_2$CoGe$_2$O$_7$} & $[110]$ & $\bar42m$ & 2.30 & 3.41 & $\sigma^{xyz}=\sigma^{yxz},\ \sigma^{zxy}$\\
0.575 & \ce{ZnFeF5(H2O)2} & $[010]$ & $mm2$  & 2.73 & 4.3 & $\sigma^{zxx},\ \sigma^{zyy},\ \sigma^{zzz},\ \sigma^{yyz},\ \sigma^{xxz}$\\
0.712 & \ce{VNb3S6} & $[100]$ & $622$ & 0 & 31.17 &$\sigma^{xyz} = -\sigma^{yxz}$\\
0.722 & \ce{Mn4Nb2O9} & $[001]$ & $m$  & 1.30 & -16.2 &\makecell{
$\sigma^{xxx}$, $\sigma^{xyy}$, $\sigma^{xzz}$, $\sigma^{xxz}$, $\sigma^{yyz}$,\\
$\sigma^{yxy}$, $\sigma^{zxx}$, $\sigma^{zyy}$, $\sigma^{zzz}$, $\sigma^{zxz}$
}\\
0.802 & \ce{CuFeS2} & $[001]$ & $\bar42m$  & 1.10 & 64.1 & $\sigma^{xyz}=\sigma^{yxz},\ \sigma^{zxy}$\\
0.823 & \ce{Sr2MnGaO5} & $[100]$ & $mm2$ & 1.27 & -6.1 & $\sigma^{zxx},\ \sigma^{zyy},\ \sigma^{zzz},\ \sigma^{yyz},\ \sigma^{xxz}$\\
0.83 & \ce{LiFeP2O7} & $\begin{array}{c}
[100]\\
+11^\circ [001]
\end{array}$ & $2$ & 2.60 & 8.78 &\makecell{
$\sigma^{xyz}$, $\sigma^{xxy}$, $\sigma^{yxx}$, $\sigma^{yyy}$,\\
$\sigma^{yzz}$, $\sigma^{yxz}$, $\sigma^{zyz}$, $\sigma^{zxy}$
}\\

\bottomrule
\end{tabular}
\end{table*}

\begin{figure}
\centering
\includegraphics[width=0.95\linewidth]{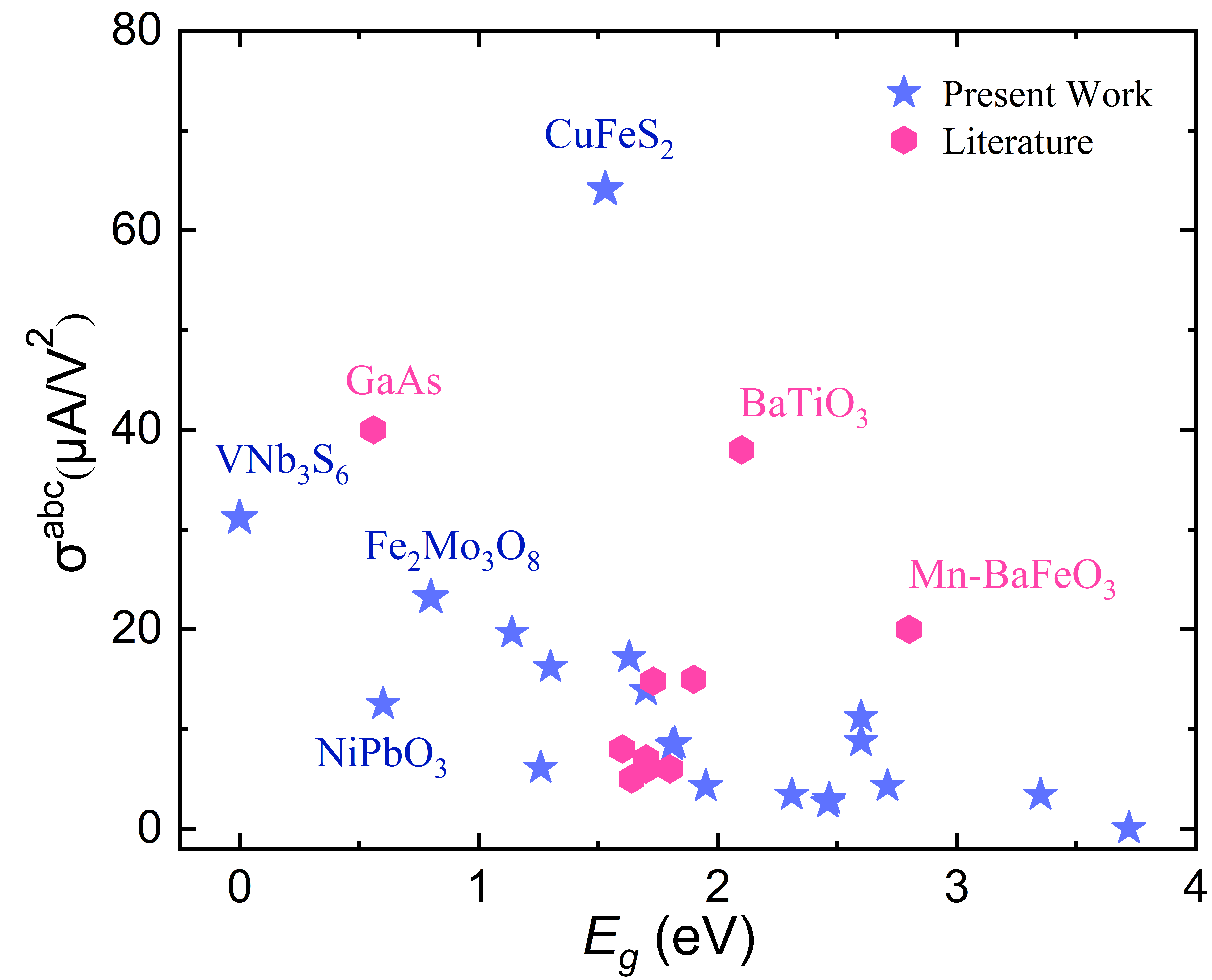}
\caption{\label{fig:all-sc} Absolute maximum shift-current conductivity as a function of band gap for non-centrosymmetric candidate materials. Blue stars denote the results from the present work, while purple diamonds represent values reported in the literature.}
\end{figure}

Beyond the Berry-curvature-driven linear responses discussed above, inversion-asymmetric ALTs provide a natural platform for exploring nonlinear effects associated with quantum geometry. 
When spatial inversion symmetry is broken, the real-space polarization structure of the crystal can couple to the altermagnetic electronic structure and generate nonlinear photocurrents. A prototypical example is the BPVE, a second-order nonlinear optical phenomenon that converts light directly into a dc current in a homogeneous crystal without requiring a \textit{p}–\textit{n} junction or interface \cite{zhao2026recent,dai2023recent}. One important microscopic mechanism of the BPVE is the shift current, which originates from the real-space shift of the electron charge center during inversion-asymmetric optical excitation, thereby generating a dc photocurrent \cite{young2012first,wang2019ferroicity,wei2026shift}.
Among the altermagnetic compounds derived from the MAGNDATA dataset, we identify 21 systems without inversion symmetry. The corresponding symmetry-allowed nonvanishing shift-current tensor components are listed in Table~\ref{tab:shiftcurrent}.

As an illustrative example of the symmetry constraints on the shift current, we consider crystals belonging to the $\bar{4}2m$ ($D_{2d}$) point group. Let $\boldsymbol{\mathcal{E}}=(\mathcal{E}_x,\mathcal{E}_y,\mathcal{E}_z)$ denote the electric-field vector. In this symmetry class, the in-plane electric-field components $(\mathcal{E}_x,\mathcal{E}_y)$ transform according to the irreducible representation $E$, whereas the out-of-plane component $\mathcal{E}_z$ transforms as $B_2$. The corresponding representation of the electric field is therefore given by $\Gamma_{\mathcal{E}} = E \oplus B_2$. The direct product of the electric-field representations can then be written as
\[
\Gamma_{\mathcal{E}} \otimes \Gamma_{\mathcal{E}} = 2A_1 \oplus A_2 \oplus B_1 \oplus B_2 \oplus 2E,
\]
while its symmetric part relevant to the shift-current response is
\[
\Gamma^{\mathrm{sym}}_{\mathcal{E}\mathcal{E}} = 2A_1 \oplus B_1 \oplus B_2 \oplus E.
\]
The photocurrent vector $\mathbf{J}=(J_x,J_y,J_z)$ transforms as $\Gamma_J = E \oplus B_2$. Accordingly, the symmetry of the shift-current response is determined by the product $\Gamma_J \otimes \Gamma^{\mathrm{sym}}_{\mathcal{E}\mathcal{E}}$. According to Neumann's principle, only terms containing the totally symmetric representation $A_1$ can give rise to nonvanishing tensor elements. As a consequence, two independent nonzero components of the shift-current tensor are allowed for the $\bar{4}2m$ point group.
Among the candidates listed in Table~\ref{tab:shiftcurrent}, \ce{Co2Mo3O8} is particularly interesting because its BPV response can be tuned through band-gap modification induced by changes in lattice parameters, as shown in Table S1. This suggests that strain engineering or other external perturbations may provide an effective route for controlling the BPV effect in altermagnetic materials \cite{kaner2020enhanced,teweng2025tunable}.

\begin{figure*}
	\centering
	\includegraphics[width=0.9\linewidth]{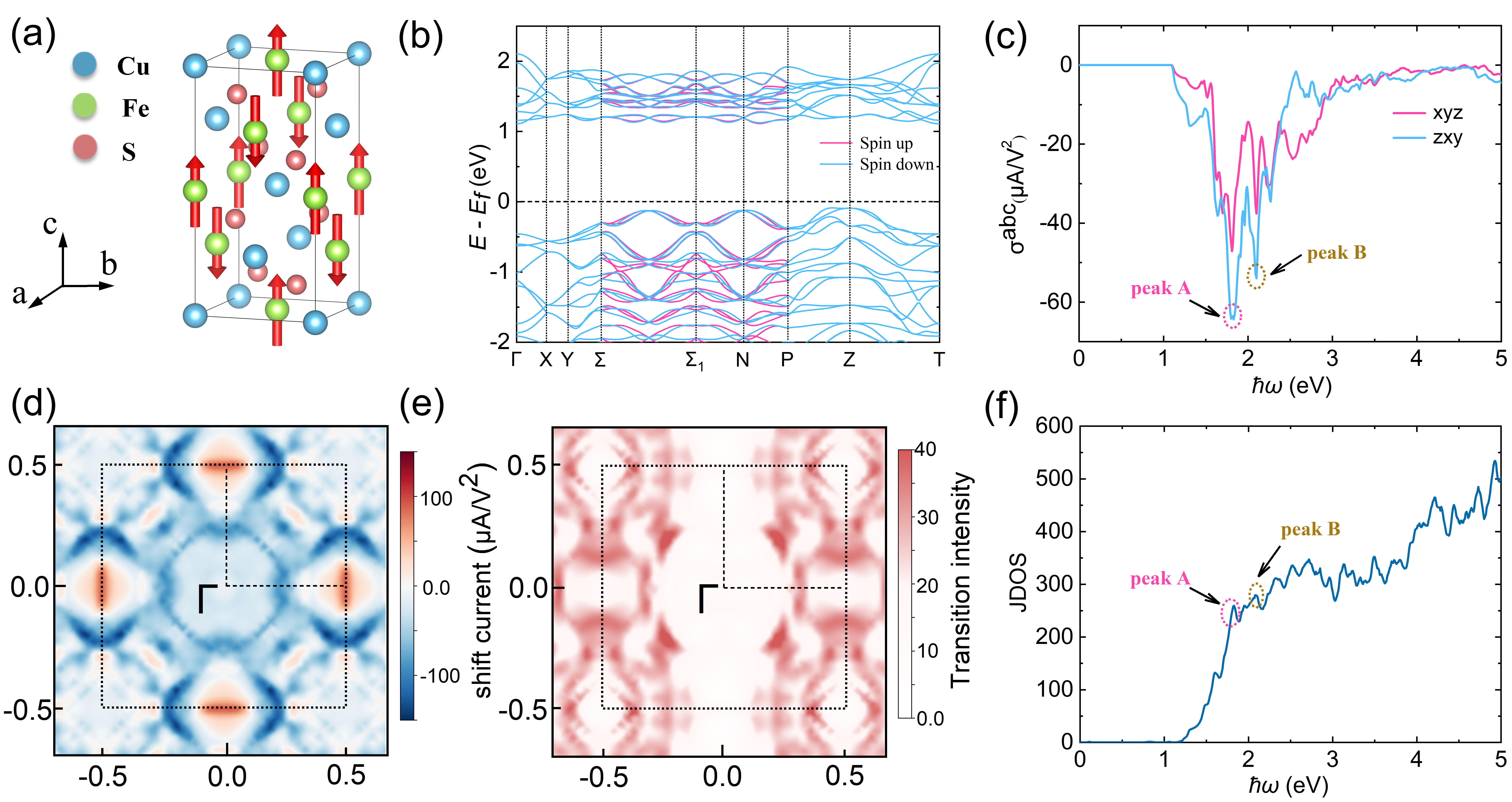}
	\caption{\label{fig:CuFeS2-sc}
(a) Crystal structure with magnetic configuration, where Cu, Fe, and S atoms are shown in blue, green, and red, respectively.
(b) Electronic band structure without SOC, showing spin splitting characteristic of altermagnetism.
(c) Frequency-dependent shift-current conductivity for the non-zero shift current components.
(d) and (e) $k$-resolved shift current distribution and transition intensity in the Brillouin zone at $\hbar\omega = 1.80~\mathrm{eV}$ for $\sigma^{xyz}$, revealing dominant contributions near high-symmetry points.
(f) Joint density of states of CuFeS$_2$.}
\end{figure*}

Figure~\ref{fig:all-sc} shows the absolute maximum shift-current conductivity as a function of band gap, compared with previously reported bulk photovoltaic materials. The calculated shift current spans a wide range across different materials, with several compounds showing particularly large responses, including \ce{VNb3S6}, \ce{CuFeS2}, and \ce{Fe2Mo3O8}. These values are comparable to or even exceed those of well-known bulk photovoltaic materials such as \ce{BaTiO3}, \ce{GaAs}, and Mn-doped \ce{BaFeO3} \cite{young2012first,feng2025high,ibanez2018ab,matsuo2017gap}, highlighting their strong potential for photovoltaic applications. Although materials with smaller band gaps generally tend to exhibit enhanced responses, the relationship between band gap and shift current is not strictly monotonic, reflecting the complex influence of electronic structure and interband transitions. In addition, \ce{VNb3S6} displays a pronounced low-frequency second-order optical response characteristic of semimetals, offering useful guidance for the design of semimetal-based terahertz photodetectors \cite{ahn2020low, okamura2026large}.

As another interesting material, \ce{CuFeS2} exhibits a giant BPVE, as shown in Figure~\ref{fig:CuFeS2-sc}. Figure~\ref{fig:CuFeS2-sc}(a) shows the crystal structure of \ce{CuFeS2}, in which Cu, Fe, and S atoms form a non-centrosymmetric lattice with finite polar displacements (indicated by red arrows) along the crystallographic $c$ axis. This collinear configuration corresponding to the magnetic space group $I\bar{4}2d$ (No.~122.333, Type I) and the spin-space-group entry No.~122.82.1.1.L \cite{chen2024enumeration}. In this symmetry setting, the two opposite-spin Fe sublattices are related by the $\bar{4}_z$ rotoinversion and associated $D_{2d}$ operations, rather than by inversion or a simple translation, which removes the global spin-degeneracy protection at generic $\mathbf{k}$ points. The corresponding spin-polarized electronic band structure Figure~\ref{fig:CuFeS2-sc}(b) shows the coexistence of spin-degenerate and spin-split regions in momentum space, characteristic of altermagnetic behavior. Because inversion symmetry is absent, a second-order photocurrent is symmetry allowed in \ce{CuFeS2}. Applying Neumann's principle to the magnetic point group $\bar{4}2m$, the symmetry-allowed shift-current tensor components reduce to $\sigma^{xyz}=\sigma^{yxz}$ and $\sigma^{zxy}$. Here, the $\bar{4}_z$ symmetry enforces the equality between $\sigma^{xyz}$ and $\sigma^{yxz}$, while the remaining tensor elements are forbidden by the twofold and mirror operations of the $D_{2d}$ group. The calculated nonzero shift current conductivity components, $\sigma^{xyz}$ and $\sigma^{zxy}$, are presented in Figure~\ref{fig:CuFeS2-sc}(c). Two pronounced peaks emerge around $\hbar\omega \approx 1.80~\mathrm{eV}$ and $\hbar\omega \approx 2.20~\mathrm{eV}$, reaching magnitudes on the order of $-64~\mu\mathrm{A/V^2}$ (peak A) and $-53~\mu\mathrm{A/V^2}$ (peak B), respectively. To further elucidate the origin of these peaks, we map the $k$-resolved shift-current distribution at $\hbar\omega = 1.80~\mathrm{eV}$ for $\sigma^{xyz}$ in Figure~\ref{fig:CuFeS2-sc}(d), which shows that the dominant contributions are concentrated in specific regions of the Brillouin zone. The close correspondence with the transition intensity distribution further indicates that the nonlinear photocurrent is governed by a limited number of hot spots. This demonstrates that the BPVE in CuFeS$_2$ is controlled by a highly non-uniform quantum-geometric texture in momentum space, as illustrated in Figure~\ref{fig:CuFeS2-sc}(e). Consistently, the joint density of states (JDOS) in Figure~\ref{fig:CuFeS2-sc}(f) exhibits clear enhancements at photon energies corresponding to peaks A and B, indicating an increased phase space for allowed interband transitions and thereby supporting the resonant character of the pronounced shift-current response.
Together, these results show that the giant BPVE arises from the interplay of symmetry-allowed tensor components, the altermagnetic band structure, and momentum-localized interband transitions, highlighting non-centrosymmetric altermagnetic materials as a promising platform for realizing large and tunable nonlinear optical responses.

\section{Conclusion}
In summary, we have established a medium-throughput first-principles framework for systematically evaluating linear and nonlinear transport responses in altermagnetic materials. By integrating DFT, automated Wannier interpolation, and symmetry analysis, we show that the quantum geometry-derived $dc-$ and $ac-$ transport properties of ALTs are governed by the interplay among crystalline/magnetic symmetry, SOC, and material-specific electronic structure.
For metallic ALTs, Berry-curvature-driven anomalous Hall and anomalous Nernst responses are strongly constrained by the magnetic space group, and that finite charge Hall signals require SOC-induced spin–lattice locking, as illustrated by \ce{VNb3S6}. For insulating ALTs, strong SOC and favorable interband transitions can produce enhanced magneto-optical Kerr responses, with \ce{CaIrO3} providing a representative case of a giant symmetry-resolved MOKE signal. For inversion-asymmetric ALTs, we identify several compounds, including \ce{CuFeS2}, \ce{Fe2Mo3O8}, and \ce{VNb3S6}, that exhibit large shift-current responses comparable to or even exceeding those of benchmark bulk photovoltaic materials.
More broadly, our work demonstrates that topological transports of ALTs are not defined by a single response channel, but by a broader landscape of symmetry-selective observables that can be activated, enhanced, and tailored through magnetic configurations, SOC, and crystalline/magnetic symmetry breaking. This provides both a practical strategy for identifying experimentally accessible fingerprints of altermagnetism and a design route toward multifunctional quantum materials for spintronic, magneto-optical, and photovoltaic applications.

\section{Methods}

All first-principles calculations were performed within DFT using the projector augmented-wave (PAW) method \cite{blochl1994projector} as implemented in VASP \cite{kresse1996efficient}. The exchange--correlation functional was treated within the generalized gradient approximation (GGA) in the Perdew--Burke--Ernzerhof (PBE) form \cite{perdew1996generalized}. SOC was included in all calculations. A $\Gamma$-centered $k$ mesh with a reciprocal-space resolution of $0.02~\mathrm{\AA^{-1}}$ was used for Brillouin-zone sampling, and the plane-wave cutoff energy was set to $550~\mathrm{eV}$. Correlated states were treated within the DFT+$U$ framework. The band gaps were evaluated using VASPKIT \cite{wang2021vaspkit}, and MLWFs were constructed from the DFT band structures using Wannier90 \cite{mostofi2014updated}. Automated Wannierization was carried out using an in-house workflow interfacing VASP and Wannier90 \cite{zhang2018high}. The resulting Wannier Hamiltonians were used for the interpolation of linear and nonlinear response functions. Linear Berry-curvature-related transport properties, including the AHC and ANC, were evaluated using WannierBerri \cite{tsirkin2021high}. The AHC was obtained through the integration of the Berry curvature across the Brillouin zone, $\Omega(\mathbf{k})$, as follows:
\begin{equation}
	\begin{split}
		& \sigma_{\alpha \beta} = -\frac{e^2}{\hbar} \int \frac{d\mathbf{k}}{(2\pi)^3} \sum f \left[\epsilon(\mathbf{k})-\mu\right] \Omega_{\alpha \beta,n} (\mathbf{k}), \\
		& \Omega_{\alpha \beta,n} (\mathbf{k}) = -2\,\mathrm{Im} \sum_{m \neq n} \frac{\braket{\mathbf{k}n|v_{\alpha}|\mathbf{k}m}\braket{\mathbf{k}m|v_{\beta}|\mathbf{k}n}}{\left[\epsilon_m(\mathbf{k})-\epsilon_n(\mathbf{k})\right]^2},
		\label{AHC+AHC-eq}
	\end{split}
\end{equation}
where $\mu$ denotes the Fermi level, and $f\left[\epsilon(\mathbf{k})-\mu\right]$ is the Fermi--Dirac distribution function. Here, $m$ and $n$ represent the occupied and empty Bloch bands, with corresponding eigenvalues $\epsilon_m(\mathbf{k})$ and $\epsilon_n(\mathbf{k})$, respectively, and $v_{\alpha}$ ($v_{\beta}$) is the velocity operator. The integration was performed using a $100\times100\times100$ mesh with WannierBerri. Similarly, the ANC is defined by
\begin{equation}
	a_{\alpha \beta} = -\frac{1}{e} \int d\epsilon \frac{\partial f}{\partial \mu}\,\sigma_{\alpha \beta}(\epsilon)\,\frac{\epsilon-\mu}{T},
	\label{ANC}
\end{equation}
where $\epsilon$ denotes a point in the energy grid, while $e$ and $T$ represent the electron charge and temperature, respectively. The integration was performed over an energy range of $[-0.5,0.5]$ eV with respect to the Fermi level at 300 K, using an energy grid consisting of 1001 points.

The Kerr rotation and ellipticity in the polar configuration, within the approximation of a semi-infinite medium, are related to the permittivity as follows:
\begin{equation}
	\theta_K(\omega) + i\eta_K(\omega) \approx 
	\frac{-\varepsilon_{ij}(\omega)}{(\varepsilon_0(\omega) - 1)\sqrt{\varepsilon_0(\omega)}} \, ,
\end{equation}
where $\varepsilon_0 = (\varepsilon_{ii} + \varepsilon_{jj})/2$. Here, $\theta_K$ and $\eta_K$ are the Kerr rotation angle and ellipticity, respectively, while $\varepsilon_{ii}$ and $\varepsilon_{ij}$ are the diagonal and off-diagonal elements of the dielectric tensor computed using VASP including SOC with a $k$-mesh density of $0.02~\mathrm{\AA^{-1}}$. This approximation is justified by the relatively thick samples ($100~\mathrm{nm}$) and their metallic behavior, which ensure that the light is absorbed within the layer and does not reach the bottom of the film. Structures with large MOKE responses were further validated by calculating the MOKE through Wannier functions. In the Wannier interpolation approach, the Kerr rotation and ellipticity are expressed in terms of the optical conductivity tensor $\sigma$ as
\begin{equation}
	\theta_K(\omega)+i\eta_K(\omega)
	=
	-\frac{\sigma_{ij}(\omega)}
	{\sigma_{0}(\omega)\sqrt{1+\dfrac{i\,\sigma_{0}(\omega)}{\varepsilon_{0}\omega}}},
    \label{eq:moke_sigma}
\end{equation}
where $\varepsilon_0$ is the vacuum permittivity, and $\sigma_0(\omega)=(\sigma_{ii}(\omega)+\sigma_{jj}(\omega))/2$. The optical conductivity tensor was calculated according to the Kubo--Greenwood theory \cite{greenwood1958boltzmann}:
\begin{equation}
	\begin{split}
		\sigma_{k}(\omega)
		&= \frac{\epsilon_{ijk}}{2}\hbar e^{2}
		\int \frac{d^{3}\mathbf{k}}{(2\pi)^{3}}
		\sum_{n \neq m}
		\frac{f_{m\mathbf{k}} - f_{n\mathbf{k}}}
		{\varepsilon_{m\mathbf{k}} - \varepsilon_{n\mathbf{k}}}
		\\
		&\quad \times
		\frac{
			\mathrm{Im}\!\left[
			\langle \psi_{n\mathbf{k}} | v_i | \psi_{m\mathbf{k}} \rangle
			\langle \psi_{m\mathbf{k}} | v_j | \psi_{n\mathbf{k}} \rangle
			\right]
		}{
			\varepsilon_{m\mathbf{k}} - \varepsilon_{n\mathbf{k}} - (\hbar\omega + i\eta)
		}.
	\end{split}
    \label{eq:opt_conductivity}
\end{equation}
with the integration performed over a $100\times100\times100$ $k$ mesh using WannierBerri. Here, $\hbar$, $e$, and $\varepsilon_{m(n)\mathbf{k}}$ represent the reduced Planck constant, positive elementary charge, and eigenenergy, respectively. The operators $\hat{v}_i$ and $\hat{v}_j$ are the $k_i$ and $k_j$ components of the velocity operator, and $|n\mathbf{k}\rangle$ denotes the eigenstate. $f(\varepsilon)$ is the Fermi-Dirac distribution funciton. The comparison between the two methods is shown in Supporting Information Sec.~3, which reveals good agreement between direct DFT and Wannier interpolation calculations.

The related circular dichroism is evaluated from the difference in optical transition probabilities induced by right- and left-circularly polarized light. Within the dipole approximation, the $k$-resolved circular dichroism is defined as
\begin{equation}
\eta(\mathbf{k}, \omega) = \frac{I_{+}(\mathbf{k}, \omega) - I_{-}(\mathbf{k}, \omega)}{I_{+}(\mathbf{k}, \omega) + I_{-}(\mathbf{k}, \omega)},
\end{equation}
where $I_{\pm}(\mathbf{k}, \omega)$ denotes the transition intensity for circular polarization $\mathbf{e}_{\pm} = (\hat{x} \pm i \hat{y})/\sqrt{2}$. The transition intensity is computed from the interband dipole matrix elements
\begin{equation}
I_{\pm}(\mathbf{k}, \omega) \propto \sum_{m,n} \left| \langle \psi_{m\mathbf{k}} \,|\, \hat{\mathbf{e}}_{\pm} \cdot \mathbf{v} \,|\, \psi_{n\mathbf{k}} \rangle \right|^2 
\, \delta(\varepsilon_{m\mathbf{k}} - \varepsilon_{n\mathbf{k}} - \hbar\omega),
\end{equation}
where $\mathbf{v}$ is the velocity operator, and the summation runs over valence ($v$) and conduction ($c$) bands. In practice, the delta function is approximated by a Lorentzian broadening with a finite lifetime parameter $\eta$. 

For second-order nonlinear optical responses, it is first necessary to clarify the different types of second-order photocurrents and their connections to band geometric quantities. In general, the total second-order photocurrent can be classified into two components: the conventional shift current and the magnetic shift current \cite{ahn2020low,jiang2025nonlinear}. In this work, we focus on the conventional shift current. The corresponding shift-current density $J_a$ can be expressed in terms of two electric-field components and the material-dependent shift-current response tensor:
\begin{equation}
	J_a = \sigma^{abc} E_b E_c
\end{equation}

The conventional shift-current tensor can be obtained using \cite{jiang2025nonlinear,gu2025ferroelectric}

\begin{equation}
	\begin{split}
		\sigma^{abc}(\omega)
		&=
		\frac{i\pi e^3}{2\hbar^2}
		\int \frac{d\mathbf{k}}{8\pi^3}
		\sum_{n,m}
		\\
		&\quad \times
		\left(
		r^{b}_{mn} r^{c}_{nm;a}
		+
		r^{c}_{mn} r^{b}_{nm;a}
		\right)
		\delta(\omega_{mn} - \omega).
	\end{split}
\end{equation}

where $a$, $b$, and $c$ denote the Cartesian directions, $r^{b}_{mn}$ marks the position-operator matrix elements, and $r^{c}_{nm;a}$ is the generalized derivative, defined as
\begin{equation}
	r^{c}_{nm;a}
	=
	\frac{\partial r^{c}_{nm}}{\partial k_a}
	-
	i\left(A^{a}_{mm} - A^{a}_{nn}\right) r^{c}_{nm},
\end{equation}
with $A^{a}_{mm}$ being the Berry connection. The conventional shift-current tensor was evaluated using WannierBerri. For systems with fewer than 250 Wannier functions, a $100\times100\times100$ $k$ mesh was employed, whereas for systems with more than 250 Wannier functions, a $60\times60\times60$ $k$ mesh was adopted. A representative $k$-mesh convergence test for \ce{CuFeS2} is shown in the Supporting Information Sec.~2.

\vspace{0.5em}
\noindent\textit{Acknowledgments.—}The authors thank the computing time provided to them on the high-performance computer Lichtenberg at the NHR Centers NHR4CES at TU Darmstadt. Fu Li acknowledges support from the China Scholarship Council. B.~Zhao and H.~Zhang acknowledge funding by the Deutsche Forschungsgemeinschaft (DFG, German Research Foundation) -- CRC 1487, ``Iron, upgraded!'' -- with project number 443703006. H.~Wang and H.~Zhang also acknowledge support from the Deutsche Forschungsgemeinschaft (DFG, German Research Foundation) under Project-ID 463184206 -- SFB 1548.

\bibliographystyle{unsrt}
\bibliography{references}

\end{document}